\documentclass{nature}
\usepackage{amsfonts,amssymb,stmaryrd,latexsym,amsmath,braket}
\usepackage{graphicx,subfigure}
\usepackage{times}
\usepackage{slashed}

\makeatletter
\let\saved@includegraphics\includegraphics
\AtBeginDocument{\let\includegraphics\saved@includegraphics}
\renewenvironment*{figure}{\@float{figure}}{\end@float}
\makeatother

\title{Projected Cooling Algorithm for Quantum Computation}
\author{Dean~Lee$^{1}$, Joey~Bonitati$^{1}$, Gabriel~Given$^{1}$, Caleb~Hicks$^{1}$, Ning~Li$^{1}$, Bing-Nan~Lu$^{1}$, Abudit~Rai$^{1}$, Avik~Sarkar$^{1}$, Jacob~Watkins$^{1}$}

\begin{document}

\maketitle 

\begin{affiliations}
\item Facility for Rare Isotope Beams and Department of Physics and
Astronomy, Michigan State University, East Lansing, MI~48824, USA
\end{affiliations}

\begin{abstract}  
  In the current era of noisy quantum devices, there is a need   for quantum algorithms that are efficient and robust against noise. Towards this end, we introduce the projected cooling algorithm for quantum computation.  The projected cooling algorithm is able to construct the localized ground state of any Hamiltonian with a translationally-invariant kinetic energy and interactions that vanish at large distances. The term ``localized'' refers to localization
in position space. The method can be viewed as the quantum analog of evaporative cooling.  We start with an initial state with support over a compact region of a large volume.  We then drive the excited quantum states to disperse and measure the remaining portion of the wave function left behind. For the nontrivial examples we consider here, the improvement over other methods is substantial.  The only additional resource required is performing the operations in a volume significantly larger than the size of the localized state. These characteristics make the projected cooling algorithm a promising tool for calculations of self-bound systems such as atomic nuclei.        
\end{abstract}

\renewcommand{\baselinestretch}{1.0}

 The quantum many-body problem is a profound challenge that pervades nearly all branches of quantum physics. While our own interests are in \textit{ab initio} methods for nuclear structure and reactions~\cite{Lu:2018bat,Elhatisari:2015iga}, the difficulties that arise at strong coupling are similar for other fields, from strongly-correlated electrons to  degenerate atomic gases, from quantum spin models to interacting relativistic quantum fields.  Quantum computing has emerged as a new computational paradigm that offers the possibility of overcoming the severe problems often faced in classical computing for the quantum many-body problem.  By allowing for arbitrary quantum superpositions of products of qubits, one can store exponentially more information than classical bits without the need for stochastic sampling. However, all of the currently available quantum computing devices suffer from short decoherence times and significant readout errors. Therefore it is necessary to develop quantum
algorithms that are efficient and robust against noise.  There are several existing methods for constructing the ground state of a quantum Hamiltonian on a quantum computer.  These include quantum phase estimation \cite{Kitaev:1995qy,Abrams:1997gk}, quantum variational methods \cite{Peruzzo:2014a,Dumitrescu:2018njn}, quantum adiabatic evolution \cite{Farhi:2000a,Farhi:2001a}, spectral comb techniques \cite{Kaplan:2017ccd}, resonance transition enhancement \cite{HWang:2017a}, coupled heat bath approaches \cite{Boykin:2002,Xu:2014}, and dissipative open-system methods \cite{Kraus:2008,Verstraete:2009}. However all of these methods have difficulties in achieving reliable accuracy for quantum Hamiltonians of interest. To address this need, in this work we introduce a method called the projected cooling algorithm. 

The projected cooling algorithm is a new approach that can be regarded as the quantum analog of evaporative
cooling.  Rather than evaporating hot gas molecules, we start with some initial state $\ket{\psi_I}$ with support over a compact region, $\rho$, and drive
excited quantum states to disperse away from $\rho$.  We then measure the remaining portion of the wave function left behind. The algorithm is able to construct the localized ground state
of any Hamiltonian with a translationally-invariant kinetic energy and interactions that go to zero at large distances. 

To illustrate, in the following we consider three different examples which we call Models 1A, 1B, and 2.  We start with a Hamiltonian  defined on a one-dimensional chain of $2L+1$ qubits at sites $n = -L,\cdots L$.  We take the vacuum to be the tensor product state where all qubits are $\ket{0}$, and from this vacuum state we can define particle excitations
in position space.  So, for example, if qubit $n$ is in the state $\ket{1}$ then we have a particle at position $n$. This is completely analogous to the spatial lattice formalism that has been used in
lattice effective field theory calculations of nuclear and cold atomic systems \cite{Lee:2008fa,Lahde:2019a}.  We are using the language of second quantization, where the number of particles equals the number of $\ket{1}$ qubits.   In this work, however, all of the examples we consider are lattice Hamiltonians that conserve particle number, just like the nuclear lattice effective field theory Hamiltonians that we hope to address in future work. For Hamiltonians
that conserve particle number, it is convenient to use the simpler language
of first quantization where the number of particles is fixed.  For this reason,
we will be discussing quantum states with a fixed number of particles and
the spatial wave functions
of such states. 

In the one-particle subspace, we let $\ket{[n]}$ be the tensor product state where qubit $n$ is $\ket{1}$ and the rest are $\ket{0}$. In the one-particle space, our Hamiltonian is defined as $H=K+V$ with $\bra{[n']}H\ket{[n]}$ equal to $K_{n',n} + V_{n}\delta_{n',n}$, 
where the kinetic energy term $K_{n',n}$ is $\delta_{n',n}-\tfrac{1}{2}\delta_{n',n+1}-\tfrac{1}{2}\delta_{n',n-1}$, and $V_n$ is the single-particle potential energy on site $n$. For the first model we consider, Model 1A, we take the interaction term to be $V_{n}=-\delta_{0,n}$, an attractive Kronecker delta function at the origin.  We  define the compact region $\rho$ to correspond to the qubits $n = -R, \cdots R$ where $R \ll L$.  We simply need
that the spatial volume of the compact region is small compared to the spatial volume of
the total system. We define $P$ to be the projection operator that projects onto the subspace where all particle excitations are contained entirely in $\rho$.  We can construct $P$ explicitly as the product of $\ket{0}\!\bra{0}$ over all qubits outside $\rho$.  Therefore $P\ket{[n]}=0$ for $|n|>R,$ and $P\ket{[n]}=\ket{[n]}$ for $|n|\le R$. 

Let $\ket{\psi_0}$ be the ground state of $H$.  For our Model 1A Hamiltonian, $\ket{\psi_0}$ is a localized bound state and is the only bound state of $H$. The term ``localized'' refers to localization
in position space. All of the other states are continuum states that extend to infinity in the limit $L \rightarrow \infty$.  Let $U(t) = e^{-iHt}$ be the time evolution operator. We are using dimensionless units for all quantities and taking $\hbar = 1$. The key result underpinning the projected cooling method is the fact that in the large volume limit $L \rightarrow \infty$, the projected time evolution operator $PU(t)P$ has a stable fixed point proportional to $P\ket{\psi_0}$.  As the time evolution operator $U(t)$ acts on $P\ket{\psi_I}$, the excited continuum states leave the compact region $\rho$. In the limit of large $t$, the only part of the wave function that remains upon projection by $P$ is from the bound state $\ket{\psi_0}$.  We are of course assuming that our initial state is not exactly orthogonal to the bound state wave function, in which case the large $t$ signal would go to zero. This fixed-point behavior will be seen for any Hamiltonian with a translationally-invariant kinetic energy, interactions that vanish at large distances, and exactly one localized bound state. If  $P\ket{\psi_I}$ has  well-defined quantum numbers associated with an exact symmetry of $H$, then the stable fixed point property applies to cases where there is exactly one localized bound state in the symmetry sector of interest.

For any two states $\ket{x}$ and $\ket{y}$, we define the normalized overlap to be $|\!\braket{x|y}\!|/\sqrt{\braket{x|x}\braket{y|y}}$.  Let $O(t)$ be the normalized overlap between $P\ket{\psi_0}$ and $PU(t)P\ket{\psi_I}$. In Fig.~1 we show $O(t)$ for five randomly chosen initial states $\ket{\psi_I}$ versus time $t$ for Model 1A.  We are choosing random initial states here in order to demonstrate that the fixed-point behavior is universal.  In this example we take
$R=5$ and $L$ large enough to prevent reflection waves returning from the boundary. We see that in all cases the normalized overlap function approaches $1$, demonstrating that $P\ket{\psi_0}$ is a stable fixed point.  We have found that the approach to the fixed point is described by a residual error that behaves as an inverse power of the projection time, $t^{-\alpha}$, but also includes sinusoidal oscillations, as can be seen in Fig.~1.  The exponent $\alpha$ as well as the frequency and amplitude of oscillations are determined by details of the system and will be discussed in a forthcoming publication.

It is helpful to make a few comments about the projection operator $P$.  The projection operator is applied only at the end of the time evolution, and   the probability of getting a nonzero signal is determined by the squared overlap between the initial state and ground state,
$\left|\braket{\psi_0|P|\psi_I}\right|^2$.  This signal efficiency is independent of the projection time and can be increased by using a better initial wave function.  For this reason a smooth initial wave function like a Gaussian wave packet
in the compact region is often a reasonable choice. 

Let us now consider any observable, $O$,  and let $O_{\rho}$ be the same observable restricted to the interior region $\rho$.  The projection operator $P$ is nothing more than the simultaneous measurement of the Pauli-$Z$ operator on each qubit exterior to $\rho$, with each qubit collapsing to $\ket{0}$ or $\ket{1}$.   We record measurement data for the expectation value of $O_{\rho}$ if and only if we find the state $\ket{0}$ for each exterior qubit.  For systems with many particle excitations, we can relax this criterion and record measurement data whenever the number of $\ket{1}$ exterior qubits is less than some small number $\delta$ times the total number of particles.

\begin{figure}
\centering
\includegraphics[width=7cm]{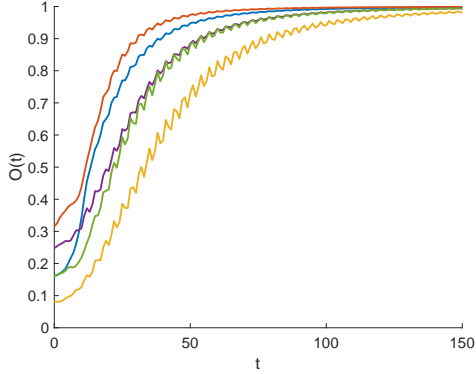}
\caption{{\bf Stable fixed point for Model
1A.}  We show the normalized overlap $O(t)$
between the exact wave function and evolved wave function over the interior
region $\rho$ versus time $t$ for Model 1A.  The plotted results correspond to five random initial states, and all approach $O(t) = 1$ for large $t$. }
\label{attractor}
\end{figure}

In order to apply projected cooling to a Hamiltonian with more than one bound state, we will need to use a time dependent Hamiltonian $H(t)$.  In the first stage of the time evolution, we let $H(t)$ be a Hamiltonian with only one bound state, but one that has good overlap with the ground state of $H$.  One simple and effective way to achieve this is to multiply the kinetic energy operator by a scale factor greater than 1 until all the bound excited states are pushed into the continuum. This augmentation
of the kinetic energy operator is also very helpful to accelerate the time evolution
at early times, as measured by the number of quantum gate operations
per qubit.
Once the time-evolved state $\ket{\psi(t)}$ is closely tracking the ground state of $H(t)$, we can then use adiabatic evolution to gradually reach the desired ground state of $H$. 

As a practical note when implementing the projected cooling algorithm, we should mention that when there is more than one localized state, there will be sinusoidal oscillations in the expectation values of operators that do not commute with the Hamiltonian.  One can therefore tune the kinetic energy scale factor until these oscillations disappear, thereby guaranteeing only one localized state.

As the next example we consider Model 1B, which is defined in exactly the same manner as Model 1A except that we take $V_{n}$ to equal $-1.6\delta_{0,n}-1.5(\delta_{2,n}+\delta_{3,n})-1.4\delta_{-2,n}$. This change is enough to produce four bound states. We first compute results using the full time evolution operator, $U(t,t-\Delta t) = e^{-iH(t)\Delta t}$. \ We then also use the Trotter approximation \cite{Trotter:1959} and break apart the Hamiltonian into pieces $H=A+B+D+V$, where $A_{n',n}$ is the off-diagonal part of $K_{n',n}$ when $\min(n',n) $ is even, $B_{n',n}$ is the off-diagonal part of $K_{n',n}$ when $\min(n',n)$ is odd, and $D_{n',n}$ is the diagonal part of $K_{n',n}$.  The Trotterized time evolution operator is then
$e^{-iA(t)\Delta t}e^{-iB(t)\Delta t}e^{-iD(t)\Delta t}e^{-iV(t)\Delta t}$.
In addition to the Trotter approximation, we also consider the effect of stochastic noise upon the time evolution.  After each $\Delta t$ time step, we multiply each component of the evolved wave function by a factor $1+z$, where $z$ is a complex Gaussian random variable centered at zero with root-mean-square values of $\varepsilon/\sqrt{2}$ for the real and imaginary parts.  

For the adiabatic evolution calculations, we start with initial Hamiltonian $H_I = V$. For the initial state we use the ground state of $H_I$, which is the point-like wave function $\ket{\psi^1_I} = \ket{[0]}$.  For the projected cooling calculations, we have more freedom and can use any initial state contained within the region $\rho$.  In addition to the point-like wave function $\ket{\psi^1_I}$, we also use the smeared wave function $\ket{\psi^2_I}$ given by $0.75\ket{[0]}+0.43\{\ket{[1]}+\ket{[-1]}\}+0.26\{\ket{[2]}+\ket{[-2]}\}$.
In Panel A of Fig.~2 we
show the normalized overlap $O(t)$ between the evolved wave function and
the exact wave function over the interior
region $\rho$  versus the number of time steps $N_t$  for Model 1B.  We take $R=5$, $L=25$, and the time step $\Delta t$ is 0.3.  This corresponds
to an interior region $\rho$ with 11 dimensions in the one-particle sector.
AE corresponds to adiabatic evolution, while PC corresponds to projected cooling.  Full evolution denotes evolution using the full time-dependent Hamiltonian for each time step, while Trotter evolution denotes the Trotter approximation.  Point initial indicates the initial state $\ket{\psi_I^1}$, while spread initial indicates the initial state $\ket{\psi_I^2}$. The quoted numerical error corresponds to the value of the parameter $\varepsilon$.  More details of these calculations can be found in Supplemental Materials. We see that  standard adiabatic evolution  has difficulties finding the ground state, achieving an overlap of no more than $0.35$.  In contrast, the projected cooling algorithm is able to achieve an overlap of at least $0.94$ in 40 time steps or less for all cases, even with the errors due to Trotter approximation and stochastic noise of size $\varepsilon = 0.05$.

\begin{figure}
\centering
\includegraphics[width=7cm]{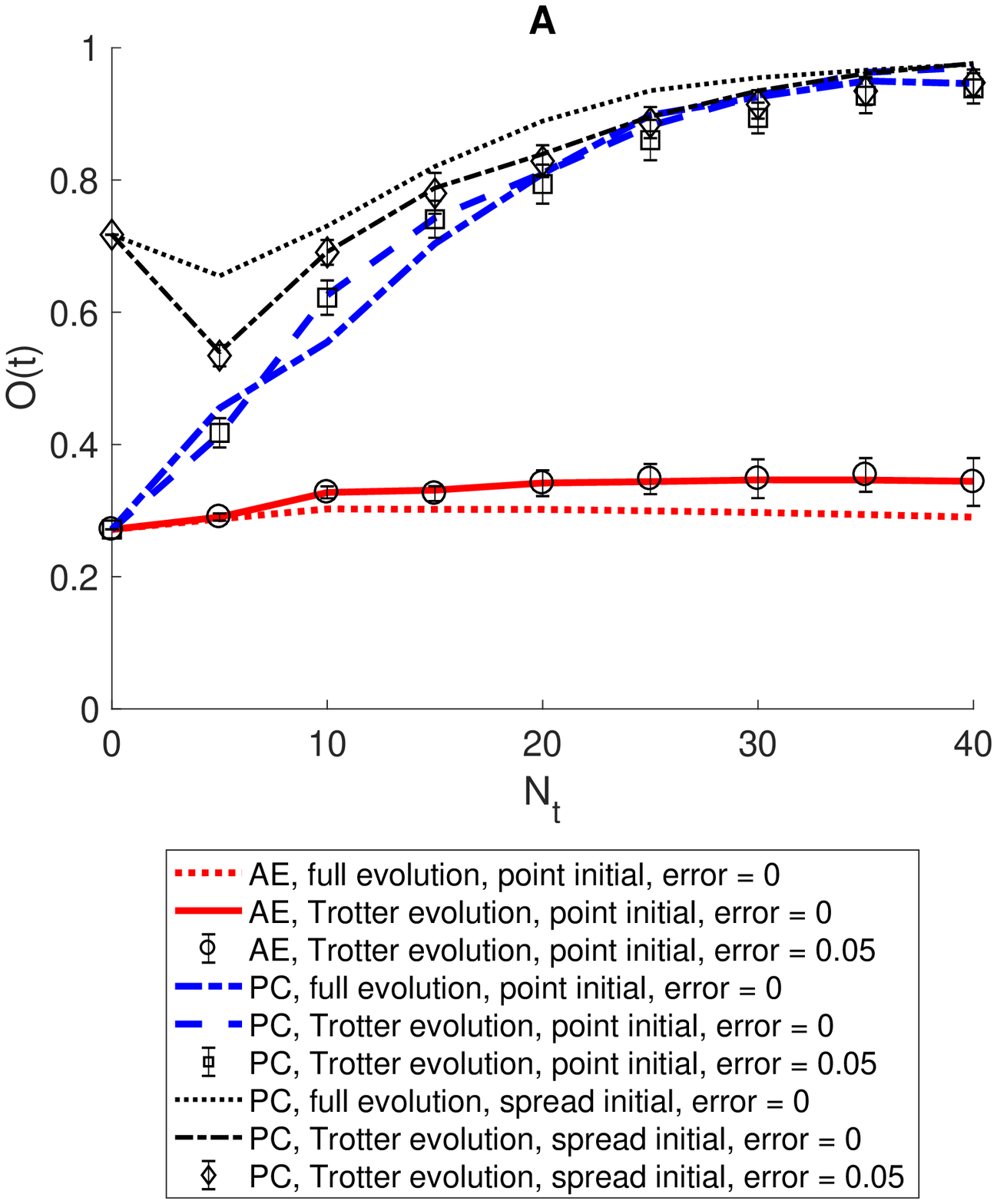}
\includegraphics[width=7cm]{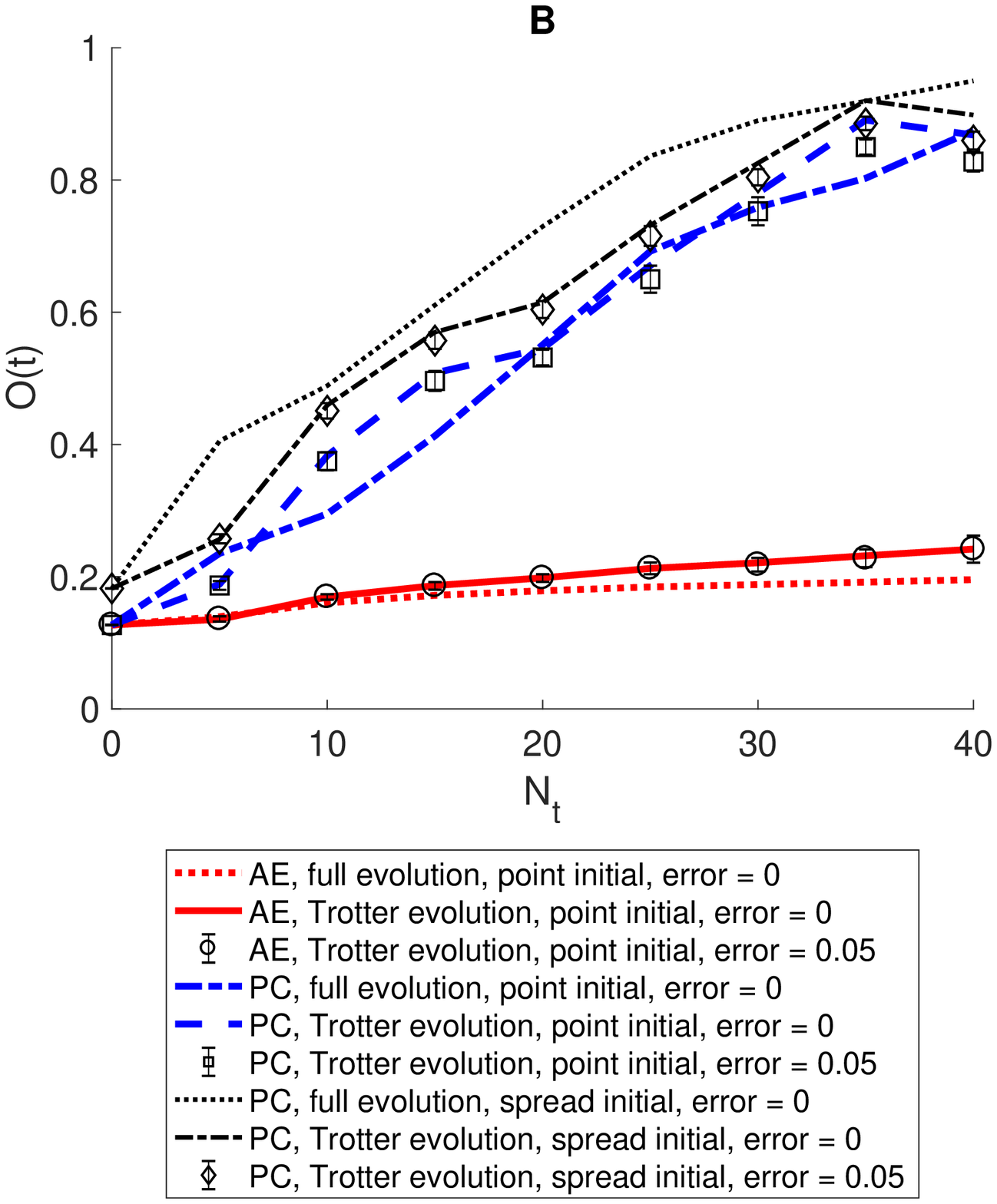}
\caption{{\bf Comparison of adiabatic evolution and projected cooling results.}  We
show the normalized overlap $O(t)$ between the evolved wave function and
the exact wave function over the interior
region $\rho$ versus the number of time steps $N_t$.  Results for Model 1B are on the left in Panel A, and results for Model 2 are on the right in Panel B.  AE is
adiabatic evolution, while PC is projected cooling.  Full evolution denotes
evolution using the full time-dependent Hamiltonian for each time step, while
Trotter evolution denotes the Trotter approximation.  Point initial indicates
the initial state $\ket{\psi_I^1}$, while spread initial indicates the initial
state $\ket{\psi_I^2}$. The quoted error corresponds to the value of the parameter $\varepsilon$.}
\label{results}
\end{figure}

For our next model, Model 2, we consider a Hamiltonian  defined on two linked one-dimensional chains of $2L+1$ qubits each, $n_1 = -L,\cdots L$ and $n_2 = -L,\cdots L$.  We again define the vacuum to be the tensor product state where all qubits are $\ket{0}$.  This time we consider the two-particle sector and define $\ket{[n_1,n_2]}$ as the tensor product state where qubit $n_1$ on the first chain is $\ket{1}$, qubit $n_2$ on
the second chain is $\ket{1}$, and all other qubits are $\ket{0}$.
  For the interior region $\rho$, we take the sites where $\max(|n_1|,|n_2|) \le R$.  We again will use the values $R=5$, $L=25$, and $\Delta t = 0.3$.  This corresponds to an interior region with 121 dimensions in the two-particle sector.  

 We refer the reader to Supplemental Materials for details of the adiabatic evolution and projected cooling calculations used for Model 2. 
 In Panel B of Fig.~2 we
show the normalized overlap $O(t)$ between the evolved wave function and
the exact wave function over the interior
region $\rho$ versus the number of time steps $N_t$  for Model 2. We see that  the standard adiabatic evolution algorithm again has difficulties finding the ground state,
achieving an overlap of no more than $0.24$.  In contrast, projected
cooling is able to achieve an overlap of at least $0.85$ in 40 time steps
or less for all cases, even with the errors due to the Trotter approximation and stochastic noise of size $\varepsilon = 0.05$.  The stochastic noise is introduced in Model 2 in exactly the same manner as in Model 1B.

In Fig.~3 we compare the evolved wave functions obtained using  adiabatic evolution and projected cooling against the exact ground state wave function.  The exact ground state wave function in the
interior region of Model 2 is shown on the left in Panel A.  The wave function
obtained via adiabatic evolution after 40 time steps is shown in the middle in Panel B,
while  the
wave function obtained
via projected cooling after 40 time steps is shown on the right in Panel C.  For these plots we used full time evolution with error $\varepsilon=0$ and the spread initial state $\ket{\psi^2_I}$ for the projected cooling calculation.
We see that the improvement obtained using projected cooling is quite significant.

The results presented here are typical of the  performance one can obtain using projected cooling.  Because of the fixed point properties of projected cooling when $H(t)$ has only one bound state, the method is flexible, efficient, and resilient against small errors.  The projected cooling algorithm is able to construct the localized ground state of any Hamiltonian with a translationally-invariant kinetic energy and interactions that go to zero at large distances, and the only additional resource required is using a volume that is significantly larger than the size of the ground state.  

The projected cooling algorithm is well suited for calculations of self-bound systems such as atomic nuclei.  For realizations on existing quantum
computing hardware, however, one should start with studies of self-bound systems in one-dimensional models first \cite{Lee:2019gtc}.  We are also working to extend the projected cooling algorithm to general Hamiltonians where there is no conservation of particle number and/or the ground state is uniform rather than localized.  It will be particularly interesting to
investigate possible connections to the phenomenon of Anderson localization
and understanding the role of the entanglement entropy between the compact
region and the external reservoir. For these cases we intend to use different Hamiltonians for the interior and exterior regions.

We note that after the appearance of our work, there was a recent application of the projected cooling algorithm to the transverse Ising model 
\cite{Gustafson:2020vqg}.  Further investigations in this direction would shed light on whether the projected cooling algorithm could also be useful for quantum annealing applications which start with a transverse field and then evolve to a classical Ising-like Hamiltonian.

\begin{figure}
\centering
\includegraphics[width=5cm]{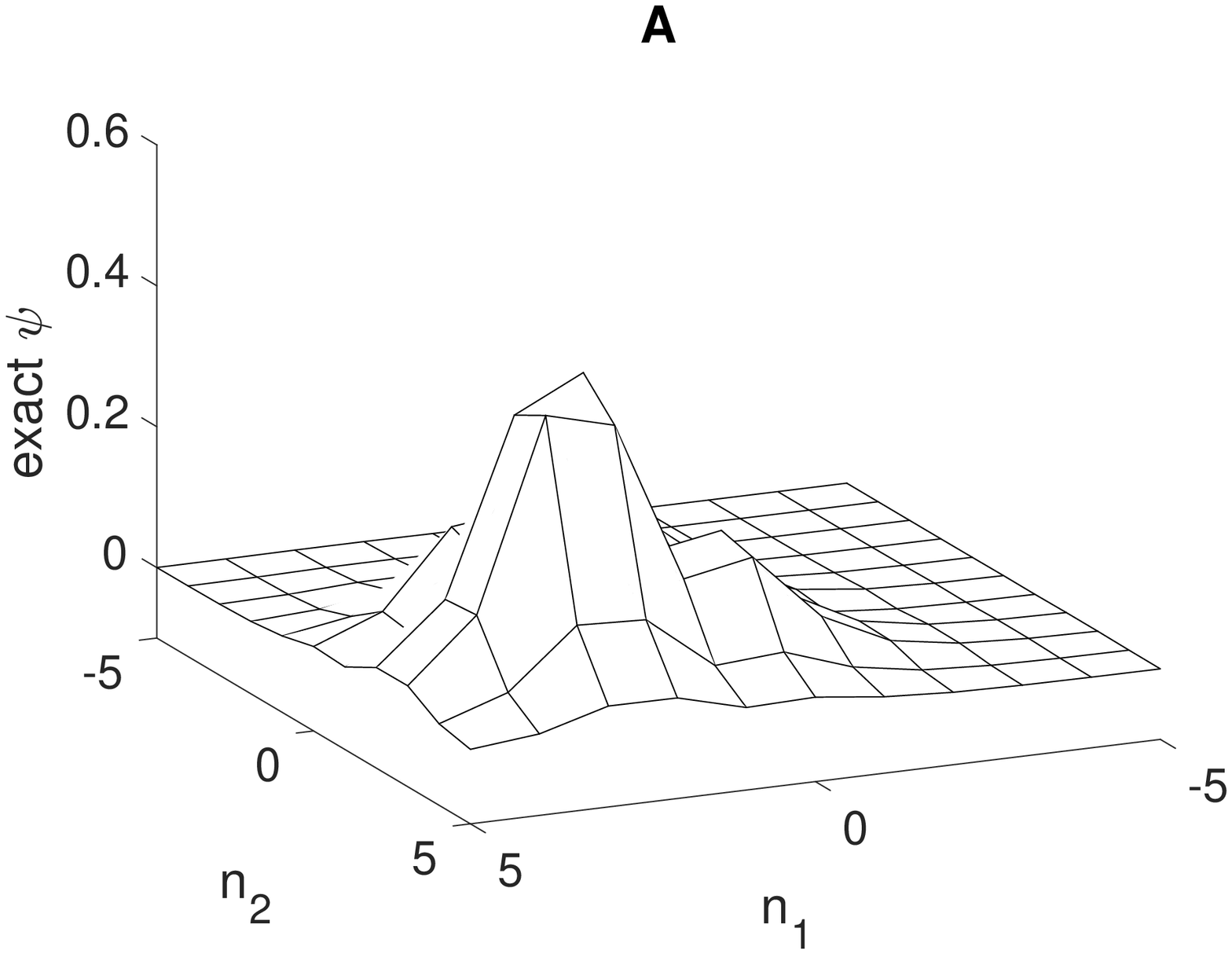}
\includegraphics[width=5cm]{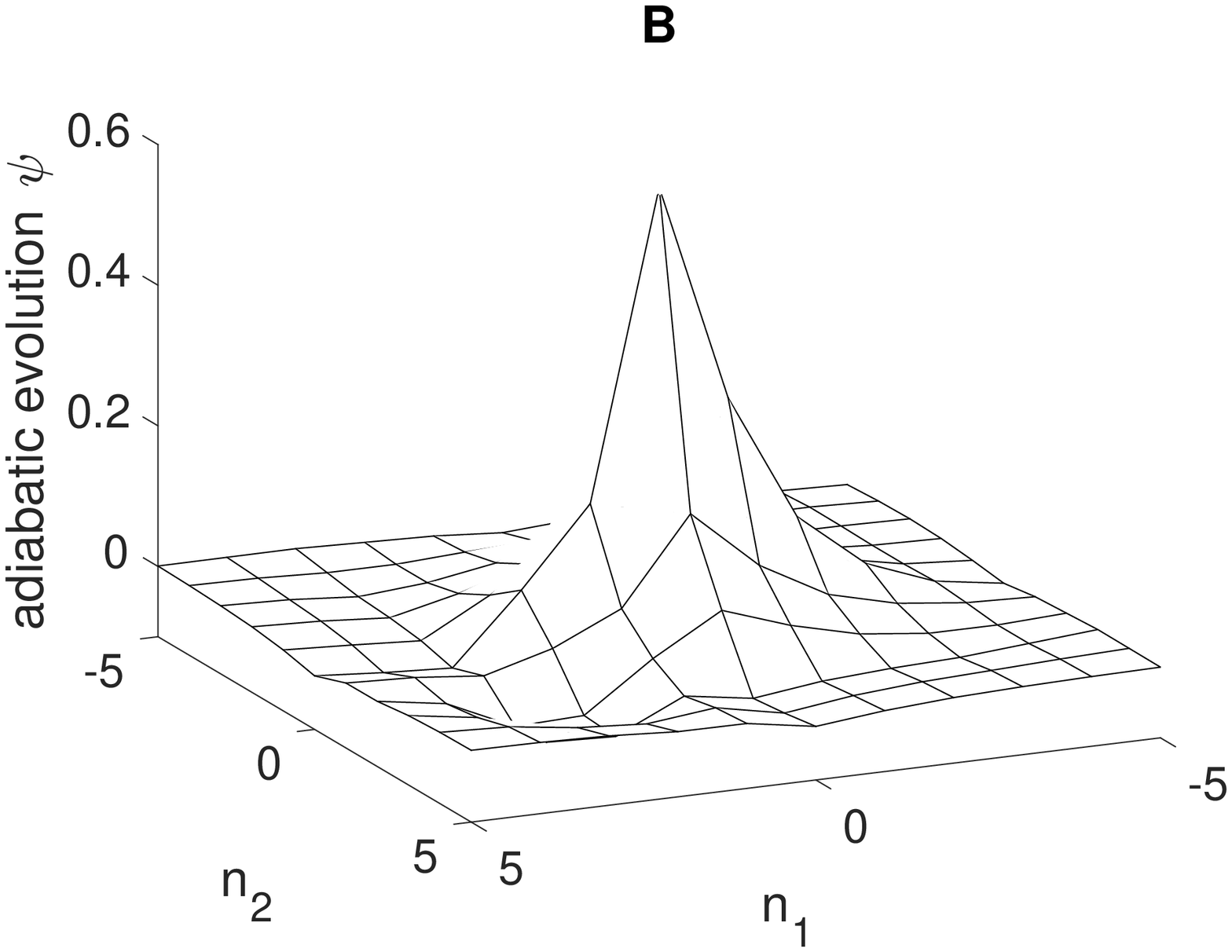}
\includegraphics[width=5cm]{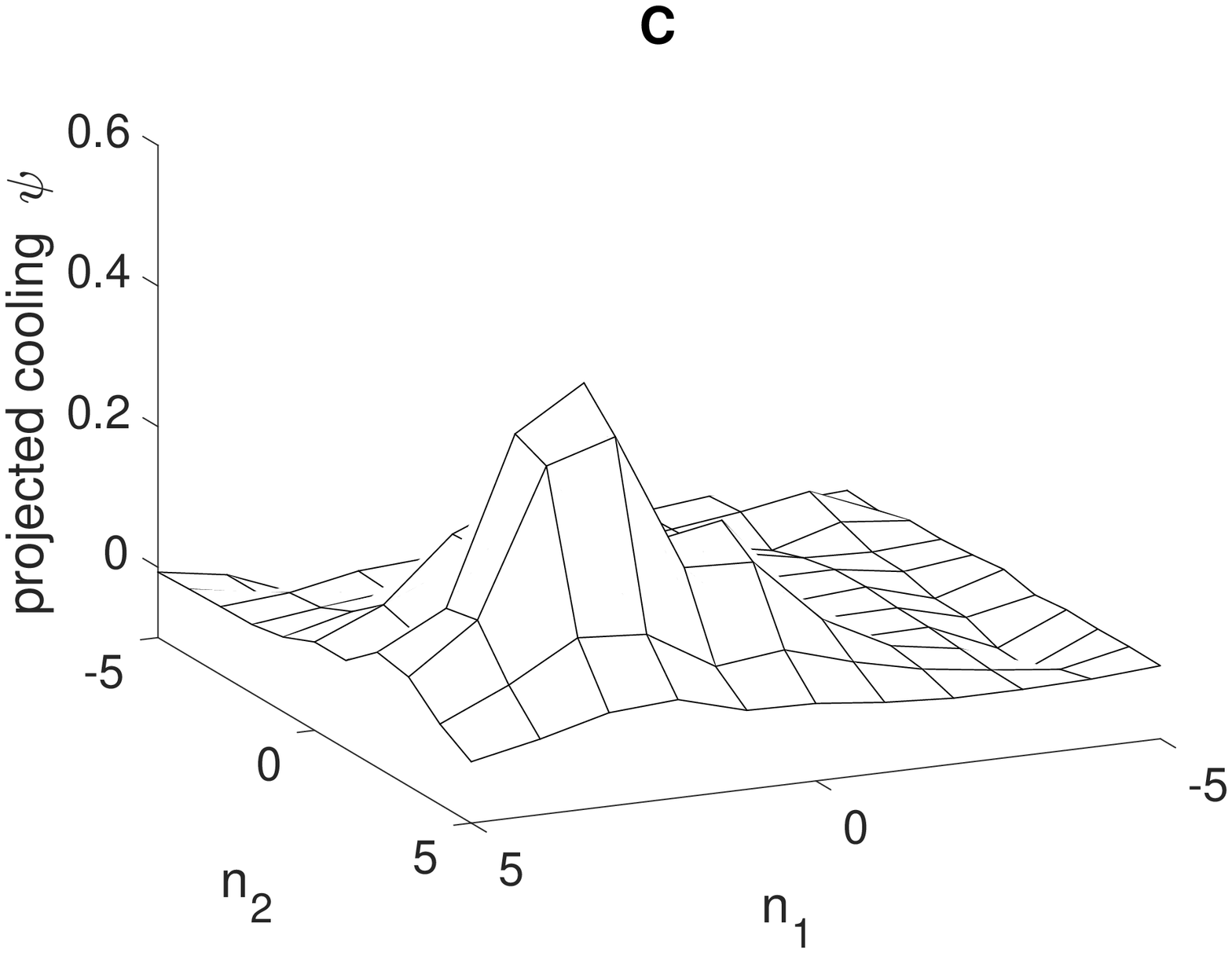}
\caption{{\bf Wave functions.} The exact ground state wave function in the
interior region of Model 2 is shown on the left in Panel A.  The wave function
obtained via adiabatic evolution after 40 time steps is shown in the middle in Panel B, while  the
wave function obtained
via projected cooling after 40 time steps is shown on the right in Panel C. We use $R=5$, $L=25$, and the time step $\Delta t$ is 0.3.}
\label{wavefunction}
\end{figure} 

\begin{addendum}
\item We thank Norman Birge and Alex Gezerlis for comments and corrections to the original draft.  We acknowledge financial support from the U.S. Department of Energy (DE-SC0018638 and DE-AC52-06NA25396).
\item[Author Contributions] D.L., J.B., N.L., B.L., and J.W. worked on algorithm and code development.  A.R. and J.W. analyzed the performance of adiabatic evolution.  D.L., C.H., G.G., and A.S. analyzed the time dependence and performance of the projected cooling algorithm.
 \item[Competing Interests] The authors declare that they have no
competing financial interests.
 \item[Correspondence] Correspondence and requests for materials
should be addressed to D.L., leed@frib.msu.edu.
\end{addendum}

\newpage

\begin{center}
\textbf{\large Supplemental Materials}
\end{center}
\setcounter{equation}{0}
\setcounter{figure}{0}
\setcounter{table}{0}
\setcounter{page}{1}
\makeatletter
\renewcommand{\theequation}{S\arabic{equation}}
\renewcommand{\thefigure}{S\arabic{figure}}

\section{Implementation of the model Hamiltonians}

We can implement Models 1A and 1B using a single one-dimensional chain of qubits.  The underlying qubit Hamiltonian has the form     
\begin{equation}
\mathcal{H} = \mathcal{D} + \mathcal{V} + \mathcal{A} + \mathcal{B},
\end{equation}
where
\begin{align}
\mathcal{A} &=- \frac{1}{4} \sum_{{\rm even}\;n} (\sigma_x^{[n+1]}\sigma_x^{[n]}
+ \sigma_y^{[n+1]}\sigma_y^{[n]}), \\
\mathcal{B} &=- \frac{1}{4} \sum_{{\rm odd}\;n} (\sigma_x^{[n+1]}\sigma_x^{[n]}
+ \sigma_y^{[n+1]}\sigma_y^{[n]}), \\
\mathcal{D}&= \frac{1}{2}\sum_n (1-\sigma_z^{[n]}), \\
\mathcal{V}&= \frac{1}{2}\sum_n V_{n} (1-\sigma_z^{[n]}).
\end{align}
Similarly, we can implement Model 2 using two linked one-dimensional chains of qubits.  The
terms in the Hamiltonian have the form
\begin{equation}
\mathcal{H} = \mathcal{A}_1 + \mathcal{B}_1 +\mathcal{A}_2 + \mathcal{B}_2 + \mathcal{D} + \mathcal{V} + \mathcal{W}, 
\end{equation}
where
\begin{align}
\mathcal{A}_1 &=- \frac{1}{4} \sum_{{\rm even}\;n_1} (\sigma_x^{[n_1+1]}\sigma_x^{[n_1]}
+ \sigma_y^{[n_1+1]}\sigma_y^{[n_1]}), \\
\mathcal{B}_1 &=- \frac{1}{4} \sum_{{\rm odd}\;n_1} (\sigma_x^{[n_1+1]}\sigma_x^{[n_1]} +\sigma_y^{[n_1+1]}\sigma_y^{[n_1]}), \\
\mathcal{A}_2 &=- \frac{1}{4} \sum_{{\rm even}\;n_2} (\sigma_x^{[n_2+1]}\sigma_x^{[n_2]}
+ \sigma_y^{[n_2+1]}\sigma_y^{[n_2]}), \\
\mathcal{B}_2 &=- \frac{1}{4} \sum_{{\rm odd}\;n_2} (\sigma_x^{[n_2+1]}\sigma_x^{[n_2]}
+ \sigma_y^{[n_2+1]}\sigma_y^{[n_2]}),\\
\mathcal{D}&= \frac{1}{4}\sum_{n_1,n_2}(1-\sigma_z^{[n_1]})(1-\sigma_z^{[n_2]}), \\
\mathcal{V}&= \frac{1}{2}\sum_{n_1} V_{n_1} (1-\sigma_z^{[n_1]}) + \frac{1}{2}\sum_{n_2} V_{n_2} (1-\sigma_z^{[n_2]}), \\
\mathcal{W}&= \frac{1}{4}\sum_{n_1,n_2} W_{n_1,n_2} (1-\sigma_z^{[n_1]})(1-\sigma_z^{[n_2]}).
\end{align}

\section{Time evolution of Model 1B}

In Model 1B our one-particle Hamiltonian
has the form $H=K+V$ with $\bra{[n']}H\ket{[n]}$ equal to $K_{n',n} + V_{n}\delta_{n',n}$, where the kinetic energy term is 
$K_{n',n} = \delta_{n',n}-\tfrac{1}{2}\delta_{n',n+1}-\tfrac{1}{2}\delta_{n',n-1}$,
and the interaction is
$V_{n}=-1.6\delta_{0,n}-1.5(\delta_{2,n}+\delta_{3,n})-1.4\delta_{-2,n}$. For the Trotter approximation we  break apart the Hamiltonian
into pieces $H=A+B+D+V$, where $A_{n',n}$ is the off-diagonal part of $K_{n',n}$
when $\min(n',n) $ is even, $B_{n',n}$ is the off-diagonal part of $K_{n',n}$
when $\min(n',n)$ is odd, and $D_{n',n}$ is the diagonal part of $K_{n',n}$.
For notational clarity when discussing   time dependent operators, we write $\bar{H}$, $\bar{K}$, $\bar{V}$, etc., to denote these  static operators.

For adiabatic evolution from initial time $t=0$ to final time $t=t_F$, we use the time-dependent Hamiltonian 
\begin{equation}
H(t) = \frac{t}{t_F} \bar{K} + \bar{V}.
\end{equation}
For projected cooling we use the time-dependent Hamiltonian \begin{equation}
H(t) = (10\bar{K}-\bar{H})\exp(-t/3.6)+\bar{H}.
\end{equation}
For the Trotterized time evolution we use the operator
\begin{equation} 
U(t,t-\Delta t) =e^{-iA(t)\Delta t}e^{-iB(t)\Delta t}e^{-iD(t)\Delta t}e^{-iV(t)\Delta
t}.
\end{equation}

\section{Time evolution of Model 2}

In Model 2 we consider a Hamiltonian  defined on two linked one-dimensional
chains of $2L+1$ qubits each, $n_1 = -L,\cdots L$ and $n_2 = -L,\cdots L$.  We define the vacuum to be the tensor product state where all qubits are $\ket{0}$.  This time we consider the two-particle sector and define $\ket{[n_1,n_2]}$ as the tensor product state where qubit $n_1$ on the first chain is $\ket{1}$, qubit $n_2$ on
the second chain is $\ket{1}$, and all other qubits are $\ket{0}$.  For the interior region $\rho$, we take the sites where $\max(|n_1|,|n_2|) \le R$.  We again will use the values $R=5$, $L=25$, and $\Delta t = 0.3$.  This corresponds to an interior region with 121 dimensions in the two-particle sector. 

The  Hamiltonian has the form $H=K+V+W$ with
\begin{equation}
\bra{[n'_1,n'_2]}H\ket{[n_1,n_2]}=K_{n'_1,n_1}\delta_{n'_2,n_2} +K_{n'_2,n_2}\delta_{n'_1,n_1}
+(V_{n_1}+V_{n_2}+W_{n_1,n_2})\delta_{n'_1,n_1}\delta_{n'_2,n_2}, 
\end{equation}
where the kinetic energy term $K_{n',n}$ is the same as in Model 1A and Model
1B.  For the interactions we take $V_{n}=-1.0\delta_{0,n} + 0.2\delta_{1,n}-0.9(\delta_{2,n}+\delta_{3,n}) -0.3\delta_{-1,n}$
for the single-particle potential energy and $W_{n_1,n_2} = -0.2\delta_{n_1,n_2}$ for the two-particle interaction.
For this model  there are
four localized bound states that remain localized in region $\rho$.   We use the point-like initial state $\ket{\psi_I^1}=\ket{[0,0]}$ for adiabatic evolution. For projected cooling we consider $\ket{\psi_I^1}$ as well as the smeared initial state 
\begin{equation}
\ket{\psi^2_I} =0.81\ket{[0,0]}+0.30\{\ket{[1,0]}+\ket{[-1,0]}+\ket{[0,1]}+\ket{[0,-1]}\}.
\end{equation}

The  Hamiltonian has the form $H=K+V+W$ with
\begin{equation}
\bra{[n'_1,n'_2]}H\ket{[n_1,n_2]}=K_{n'_1,n_1}\delta_{n'_2,n_2} +K_{n'_2,n_2}\delta_{n'_1,n_1}
+(V_{n_1}+V_{n_2}+W_{n_1,n_2})\delta_{n'_1,n_1}\delta_{n'_2,n_2}, 
\end{equation}
where the kinetic energy term is  $K_{n',n} = \delta_{n',n}-\tfrac{1}{2}\delta_{n',n+1}-\tfrac{1}{2}\delta_{n',n-1}$.
For the interactions we take 
$V_{n} = -1.0\delta_{0,n} +0.2\delta_{1,n}-0.9(\delta_{2,n} + \delta_{3,n})
-0.3\delta_{-1,n}$
and $W_{n_1,n_2} = -0.2\delta_{n_1,n_2}$.
 For the Trotter approximation we  break apart the Hamiltonian
into pieces 
\begin{equation}
H=A^{[1]}+B^{[1]}+A^{[2]}+B^{[2]}+D+V+W,
\end{equation}
where $A^{[i]}_{n'_i,n_i}$ is the off-diagonal part of the kinetic energy for particle $i$ when $\min(n'_i,n_i) $ is even, $B^{[i]}_{n'_i,n_i}$ is the off-diagonal part of the kinetic energy when $\min(n'_i,n_i)$ is odd, and $D$ is the diagonal part of the kinetic energy.  When discussing time dependent operators, we write
$\bar{H}$, $\bar{K}$, $\bar{V}$, etc., to denote these  static operators.

For the adiabatic evolution from initial time $t=0$ to final time $t=t_F$, we use the time-dependent Hamiltonian 
\begin{equation}
H(t) = \frac{t}{t_F} \bar{K} + \bar{V} + \bar{W}.
\end{equation}
For the projected cooling  we use the time-dependent Hamiltonian
\begin{equation}
H(t) = (10\bar{K}-\bar{H})\exp(-t/3.6)+\bar{H}.
\end{equation}For the Trotterized time evolution we use the operator
\begin{align} 
U(t,&t-\Delta t) = \nonumber \\
&e^{-iA^{[1]}(t)\Delta t}e^{-iB^{[1]}(t)\Delta t}
e^{-iA^{[2]}(t)\Delta t}e^{-iB^{[2]}(t)\Delta t}
e^{-iD(t)\Delta t}e^{-iV(t)\Delta t}e^{-iW(t)\Delta t}.
\end{align}

\bibliography{References}

\begin{thebibliography}{10}
\expandafter\ifx\csname url\endcsname\relax
  \def\url#1{\texttt{#1}}\fi
\expandafter\ifx\csname urlprefix\endcsname\relax\def\urlprefix{URL }\fi
\providecommand{\bibinfo}[2]{#2}
\providecommand{\eprint}[2][]{\url{#2}}

\bibitem{Lu:2018bat}
\bibinfo{author}{Lu, B.-N.} \emph{et~al.}
\newblock \bibinfo{title}{{Essential elements for nuclear binding}}.
\newblock \emph{\bibinfo{journal}{Phys. Lett.}}
  \textbf{\bibinfo{volume}{B797}}, \bibinfo{pages}{134863}
  (\bibinfo{year}{2019}).
\newblock \eprint{1812.10928}.

\bibitem{Elhatisari:2015iga}
\bibinfo{author}{Elhatisari, S.} \emph{et~al.}
\newblock \bibinfo{title}{{Ab initio alpha-alpha scattering}}.
\newblock \emph{\bibinfo{journal}{Nature}} \textbf{\bibinfo{volume}{528}},
  \bibinfo{pages}{111} (\bibinfo{year}{2015}).
\newblock \eprint{1506.03513}.

\bibitem{Kitaev:1995qy}
\bibinfo{author}{Kitaev, A.~{\relax Yu}.}
\newblock \bibinfo{title}{{Quantum measurements and the Abelian stabilizer
  problem}}  (\bibinfo{year}{1995}).
\newblock \eprint{quant-ph/9511026}.

\bibitem{Abrams:1997gk}
\bibinfo{author}{Abrams, D.~S.} \& \bibinfo{author}{Lloyd, S.}
\newblock \bibinfo{title}{{Simulation of many body Fermi systems on a universal
  quantum computer}}.
\newblock \emph{\bibinfo{journal}{Phys. Rev. Lett.}}
  \textbf{\bibinfo{volume}{79}}, \bibinfo{pages}{2586--2589}
  (\bibinfo{year}{1997}).
\newblock \eprint{quant-ph/9703054}.

\bibitem{Peruzzo:2014a}
\bibinfo{author}{{Peruzzo}, A.} \emph{et~al.}
\newblock \bibinfo{title}{{A variational eigenvalue solver on a photonic
  quantum processor}}.
\newblock \emph{\bibinfo{journal}{Nature Communications}}
  \textbf{\bibinfo{volume}{5}}, \bibinfo{pages}{4213} (\bibinfo{year}{2014}).
\newblock \eprint{1304.3061}.

\bibitem{Dumitrescu:2018njn}
\bibinfo{author}{Dumitrescu, E.~F.} \emph{et~al.}
\newblock \bibinfo{title}{{Cloud Quantum Computing of an Atomic Nucleus}}.
\newblock \emph{\bibinfo{journal}{Phys. Rev. Lett.}}
  \textbf{\bibinfo{volume}{120}}, \bibinfo{pages}{210501}
  (\bibinfo{year}{2018}).
\newblock \eprint{1801.03897}.

\bibitem{Farhi:2000a}
\bibinfo{author}{{Farhi}, E.}, \bibinfo{author}{{Goldstone}, J.},
  \bibinfo{author}{{Gutmann}, S.} \& \bibinfo{author}{{Sipser}, M.}
\newblock \bibinfo{title}{Quantum computation by adiabatic evolution}
  (\bibinfo{year}{2000}).
\newblock \eprint{quant-ph/0001106}.

\bibitem{Farhi:2001a}
\bibinfo{author}{Farhi, E.} \emph{et~al.}
\newblock \bibinfo{title}{A quantum adiabatic evolution algorithm applied to
  random instances of an np-complete problem}.
\newblock \emph{\bibinfo{journal}{Science}} \textbf{\bibinfo{volume}{292}},
  \bibinfo{pages}{472--475} (\bibinfo{year}{2001}).

\bibitem{Kaplan:2017ccd}
\bibinfo{author}{Kaplan, D.~B.}, \bibinfo{author}{Klco, N.} \&
  \bibinfo{author}{Roggero, A.}
\newblock \bibinfo{title}{{Ground States via Spectral Combing on a Quantum
  Computer}}  (\bibinfo{year}{2017}).
\newblock \eprint{1709.08250}.

\bibitem{HWang:2017a}
\bibinfo{author}{Wang, H.}
\newblock \bibinfo{title}{Quantum algorithm for preparing the ground state of a
  system via resonance transition}.
\newblock \emph{\bibinfo{journal}{Sci. Rep.}} \textbf{\bibinfo{volume}{7}},
  \bibinfo{pages}{16342} (\bibinfo{year}{2017}).

\bibitem{Boykin:2002}
\bibinfo{author}{Boykin, P.~O.}, \bibinfo{author}{Mor, T.},
  \bibinfo{author}{Roychowdhury, V.}, \bibinfo{author}{Vatan, F.} \&
  \bibinfo{author}{R., V.}
\newblock \bibinfo{title}{{Algorithmic cooling and scalable NMR quantum
  computers}}.
\newblock \emph{\bibinfo{journal}{Proc. Natl. Acad. Sci.}}
  \textbf{\bibinfo{volume}{99}}, \bibinfo{pages}{3388} (\bibinfo{year}{2002}).

\bibitem{Xu:2014}
\bibinfo{author}{{Xu}, J.-S.} \emph{et~al.}
\newblock \bibinfo{title}{{Demon-like algorithmic quantum cooling and its
  realization with quantum optics}}.
\newblock \emph{\bibinfo{journal}{Nature Photonics}}
  \textbf{\bibinfo{volume}{8}}, \bibinfo{pages}{113} (\bibinfo{year}{2014}).

\bibitem{Kraus:2008}
\bibinfo{author}{{Kraus}, B.} \emph{et~al.}
\newblock \bibinfo{title}{{Preparation of entangled states by quantum Markov
  processes}}.
\newblock \emph{\bibinfo{journal}{Phys. Rev. A}} \textbf{\bibinfo{volume}{78}},
  \bibinfo{pages}{042307} (\bibinfo{year}{2008}).

\bibitem{Verstraete:2009}
\bibinfo{author}{Verstraete, F.}, \bibinfo{author}{Wolf, M.~M.} \&
  \bibinfo{author}{Ignacio~Cirac, J.}
\newblock \bibinfo{title}{{Quantum computation and quantum-state engineering
  driven by dissipation}}.
\newblock \emph{\bibinfo{journal}{Nature Physics}} .

\bibitem{Lee:2008fa}
\bibinfo{author}{Lee, D.}
\newblock \bibinfo{title}{{Lattice simulations for few- and many-body
  systems}}.
\newblock \emph{\bibinfo{journal}{Prog. Part. Nucl. Phys.}}
  \textbf{\bibinfo{volume}{63}}, \bibinfo{pages}{117--154}
  (\bibinfo{year}{2009}).
\newblock \eprint{0804.3501}.

\bibitem{Lahde:2019a}
\bibinfo{author}{L{\"a}hde, T.} \& \bibinfo{author}{Mei{\ss}ner, U.-G.}
\newblock \emph{\bibinfo{title}{Nuclear Lattice Effective Field Theory}}
  (\bibinfo{publisher}{Springer International Publishing},
  \bibinfo{year}{2019}).

\bibitem{Trotter:1959}
\bibinfo{author}{Trotter, H.~F.}
\newblock \bibinfo{title}{On the product of semi-groups of operators}.
\newblock \emph{\bibinfo{journal}{Proc. Amer. Math. Soc.}}
  \textbf{\bibinfo{volume}{10}}, \bibinfo{pages}{545} (\bibinfo{year}{1959}).

\bibitem{Lee:2019gtc}
\bibinfo{author}{Lee, D.} \emph{et~al.}
\newblock \bibinfo{title}{{Time fractals and discrete scale invariance with
  trapped ions}}.
\newblock \emph{\bibinfo{journal}{Phys. Rev.}} \textbf{\bibinfo{volume}{A100}},
  \bibinfo{pages}{011403} (\bibinfo{year}{2019}).
\newblock \eprint{1901.01661}.

\bibitem{Gustafson:2020vqg}
\bibinfo{author}{Gustafson, E.}
\newblock \bibinfo{title}{{Projective Cooling for the transverse Ising model}}.
\newblock \emph{\bibinfo{journal}{Phys. Rev. D}}
  \textbf{\bibinfo{volume}{101}}, \bibinfo{pages}{071504}
  (\bibinfo{year}{2020}).
\newblock \eprint{2002.06222}.

\end{thebibliography}

\end{document}